\documentclass[aps,prl,reprint,superscriptaddress,floatfix]{revtex4-2}
\usepackage[colorlinks=true,allcolors=blue]{hyperref} 
\usepackage{graphicx} 
\usepackage{bm}
\usepackage{amsmath, amssymb}
\usepackage{braket}
\usepackage{appendix}

\newcommand{\Figref}[1]{Fig.~\ref{#1}}

\newcommand{\Eqref}[1]{Eq.~(\ref{#1})}

\begin{document}

\title{Viscotaxis of chiral microswimmer in viscosity gradients}

\author{Takuya Kobayashi}
\email{kobayashi@cheme.kyoto-u.ac.jp}
\affiliation{
Department of Chemical Engineering, Kyoto University, Kyoto 615-8510, Japan
}
\author{Ryoichi Yamamoto}
\email{ryoichi@cheme.kyoto-u.ac.jp}
\affiliation{
Department of Chemical Engineering, Kyoto University, Kyoto 615-8510, Japan
}

\date{\today}
 
\begin{abstract}
Microswimmers display an intriguing ability to navigate through fluids with spatially varying viscosity, a behavior known as viscotaxis, which plays a crucial role in guiding their motion. In this study, we reveal that the orientation dynamics of chiral squirmers in fluids with uniform viscosity gradients can be elegantly captured using the Landau–Lifshitz–Gilbert equations, originally developed for spin systems. Remarkably, we discover that chiral swimmers demonstrate negative viscotaxis, tracing spiral trajectories as they move. Specifically, a chiral squirmer with a misaligned source dipole and rotlet dipole exhibits a steady-state spiral motion—a stark contrast to the linear behavior observed when the dipoles are aligned. This work provides fresh insights into the intricate interplay between microswimmer dynamics and fluid properties.
\end{abstract}

\maketitle

{\it Introduction.}---Microswimmers navigate through inhomogeneous environments characterized by gradients of fields such as light intensity~\cite{Jekely2008-bx, Bennett2015-rw, Giometto2015-qb, Lozano2016-xj, Dai2016-iq}, magnetic fields~\cite{Klumpp2016-tc, Rupprecht2016-ff, Waisbord2016-nr}, temperature~\cite{Bahat2003-sl, Li2013-tc, Bickel2014-io}, and gravitational potential~\cite{Campbell2013-fv, Ten_Hagen2014-he, Wolff2013-kq, Campbell2017-ii, Shaik2024-xt}. In such environments, microswimmers respond to these spatial gradients through a phenomenon known as physicotaxis, which plays a crucial role in processes such as optimizing nutrient uptake, avoiding toxins, and guiding sperm cells toward eggs~\cite{Eisenbach2006-vw}.
The mechanisms of taxis also offer significant potential for designing microrobots with biomedical applications, including micro-cargo transport~\cite{Baraban2011-dq, Boymelgreen2018-bh}, targeted drug delivery~\cite{Srivastava2016-xt, Bhuyan2017-rc, Bunea2020-fv}, and microsurgery~\cite{Nelson2010-uz, Srivastava2016-xt, Bunea2019-yi, Bunea2020-fv}. However, the physical mechanisms governing microswimmer behavior in fluids with spatial gradients remain largely unexplored.

We focus on viscosity gradients found in various biological systems, such as human viscera~\cite{Wheeler2019-ae} and biofilms~\cite{Hall-Stoodley2004-uw}, which play crucial roles in mucosal barrier function~\cite{Swidsinski2007-jr} and influence cell motility~\cite{Wheeler2019-ae}.
Experimental studies reveal diverse responses to viscosity gradients. For instance, {\it Leptospira}\cite{Kaiser1975-nl, Petrino1978-wh, Takabe2017-tp} and {\it Spiroplasma}\cite{Daniels1980-ue} exhibit positive viscotaxis, moving up viscosity gradients, while {\it E. coli} demonstrates negative viscotaxis, moving down these gradients~\cite{Sherman1982-ur}. Additionally, {\it Chlamydomonas reinhardtii} displays unique behaviors: it shows negative viscotaxis in sufficiently strong gradients, accumulates in high-viscosity regions under weak gradients~\cite{Stehnach2021-nv}, and scatters away from interfaces when crossing from low-viscosity to high-viscosity regions~\cite{Coppola2021-iu}.

Theoretical studies have also explored both positive and negative viscotaxis~\cite{Liebchen2018-yb, Datt2019-jv, Shaik2021-ha, Gong2023-xt, Gong2024-no, Gong2024-yi}. For instance, models of {\it Leptospira} and {\it Spiroplasma} demonstrate positive viscotaxis~\cite{Liebchen2018-yb}. The velocities of achiral squirmers, characterized by their first two polar modes, were analytically derived by Datt~\cite{Datt2019-jv} as follows:
\begin{subequations}
    \begin{align}
    \bm{U} &= \bm{U}_N - \frac{aB_2}{5}(\bm{1} - 3\mathbf{pp})\cdot\bm{\nabla}\left(\frac{\eta}{\eta_0}\right),\label{eq:vel_datt}\\
    \bm{\Omega} &= -\frac{1}{2}\bm{U}_N\times \bm{\nabla}\left(\frac{\eta}{\eta_0}\right),
\end{align}
\end{subequations}
where $\bm{U}_N = (2 / 3)B_1\mathbf{p}$ is the velocity in uniform Newtonian fluids (with $\mathbf{p}$ representing the swimming axis), $a$ is the swimmer’s radius, $\eta_0$ is the characteristic viscosity, and $\bm{\nabla}\eta$ is the viscosity gradient.
Squirmers exhibit negative viscotaxis, with their velocities depending on their swimming mode. Interestingly, squirmers with only the $B_2$ mode can swim in linear viscosity gradients even without propulsion in uniform Newtonian fluids.

Microswimmers often utilize chirality to convert rotational motion into translational propulsion. While experimental investigations have explored chiral microswimmers like {\it E. coli} cells, theoretical studies typically focus on microswimmers without chirality, leaving chiral swimmers less understood. In this Letter, we model chiral microswimmers such as {\it E. coli} using a sphere with rotlet dipoles to investigate the effects of chirality on locomotion. Previous studies have shown that chirality, represented by the rotlet dipole, can significantly influence swimmer behavior. For instance, chirality enhances swimming velocity in uniform viscoelastic fluids~\cite{Binagia2020-dn, Housiadas2021-qe, Kobayashi2023-ad, Kobayashi2024-tn}, while reducing it in uniform shear-thinning fluids~\cite{Nganguia2020-ss}. Additionally, chiral spherical microswimmers exhibit circular trajectories on a flat wall~\cite{Fadda2020-ve}, consistent with experimental observations of {\it E. coli}~\cite{Lauga2006-mv}. Exploring chiral swimmer locomotion in viscosity gradients could offer valuable insights into cell motility in complex biological systems, but this remains an open question.

\begin{figure}[tb!]
    \centering
    \includegraphics[width=\linewidth]{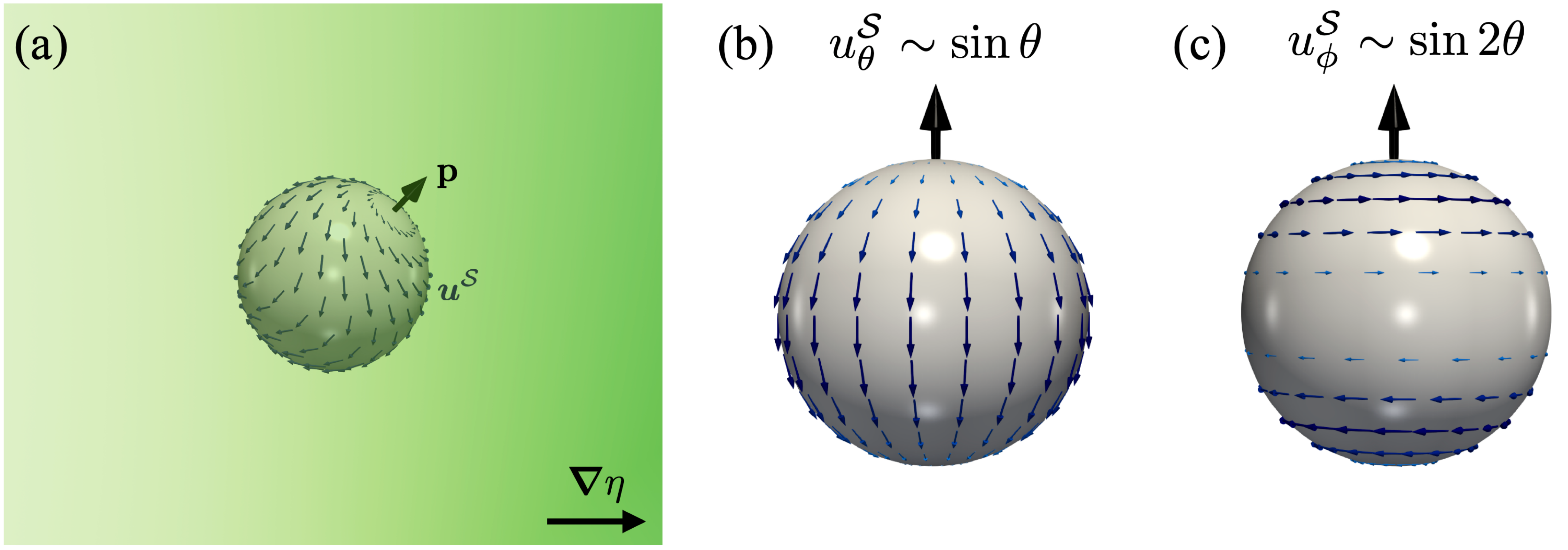}
    \caption{
    Schematic representations of (a) the motion of a squirmer with orientation $\mathbf{p}$ in fluids exhibiting viscosity gradients, and (b) and (c) the surface slip velocities $\bm{u}^\mathcal{S}$ corresponding to the $B_1$ and $C_2$ modes, respectively.
    }
    \label{fig1}
\end{figure}

\begin{figure}[tb!]
    \centering
    \includegraphics[width=\linewidth]{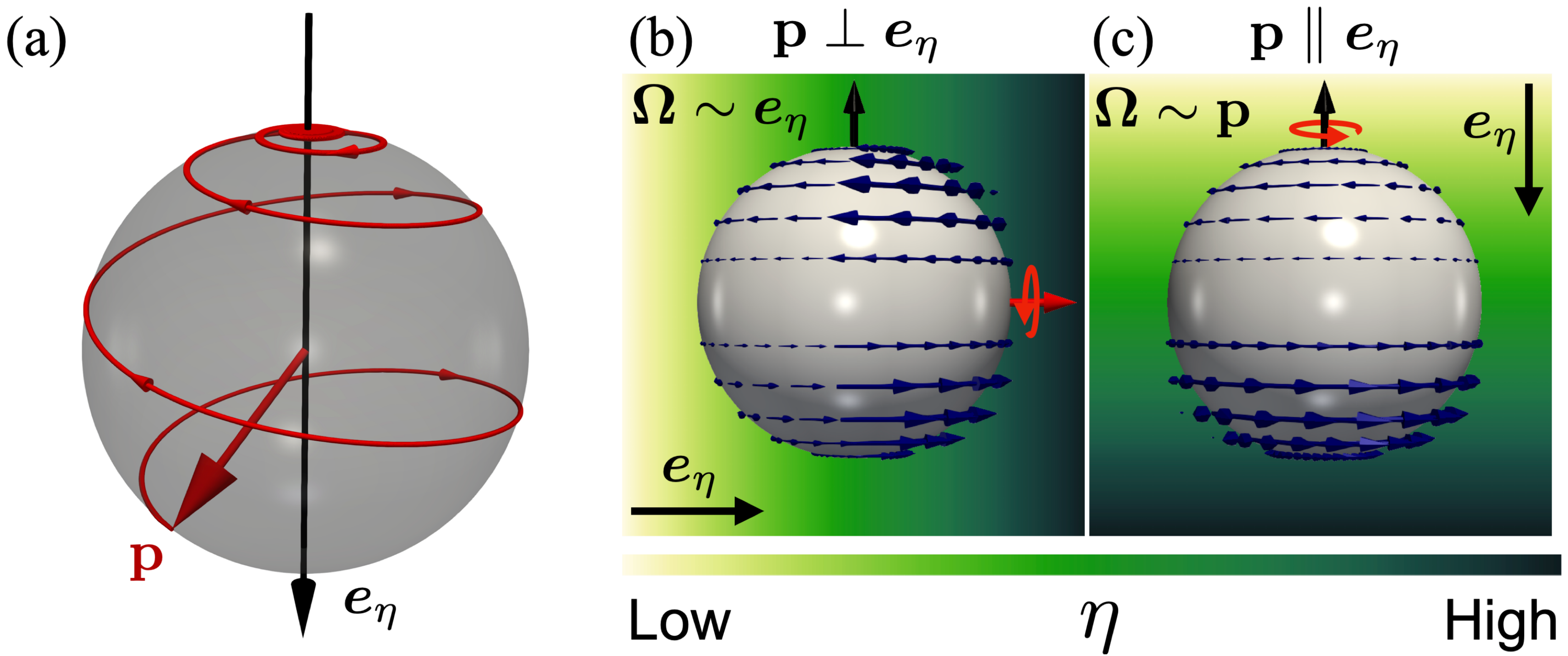}
    \caption{
    (a) Time evolution of the orientation $\mathbf{p}$ of the chiral swimmer.
    (b) and (c) Schematic representations of the mechanism driving the rotation, caused by the coupling between chirality and the viscosity gradient, for the (b) perpendicular ($\mathbf{p} \perp \bm{e}\eta$) and (c) parallel ($\mathbf{p} \parallel \bm{e}\eta$) cases.    
    }
    \label{fig2}
\end{figure}

\begin{figure}[tb!]
    \centering
    \includegraphics[width=\linewidth]{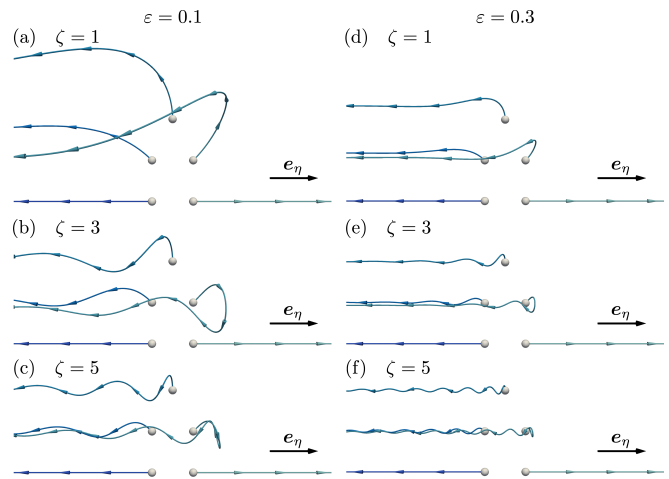}
    \caption{
    Trajectories of chiral squirmers for various initial polar angles $\vartheta(0)$ and chiral parameters $\zeta = C_2 / B_1$: (a), (d) $\zeta = 1$; (b), (e) $\zeta = 3$; and (c), (f) $\zeta = 5$. The viscosity gradient strengths are set to $\varepsilon = 0.1$ in (a)-(c) and $\varepsilon = 0.3$ in (d)-(f). The initial azimuthal angle is fixed at $\varphi(0) = 0$. When the swimmer’s orientation is perfectly parallel or antiparallel to the viscosity gradient axis $\bm{e}_\eta$, it follows a straight trajectory. However, if the orientation deviates from the axis, the swimmer follows spiral trajectories, exhibiting negative viscotaxis.
    }
    \label{fig3}
\end{figure}

\begin{figure}[tb!]
    \centering
    \includegraphics[width=\linewidth]{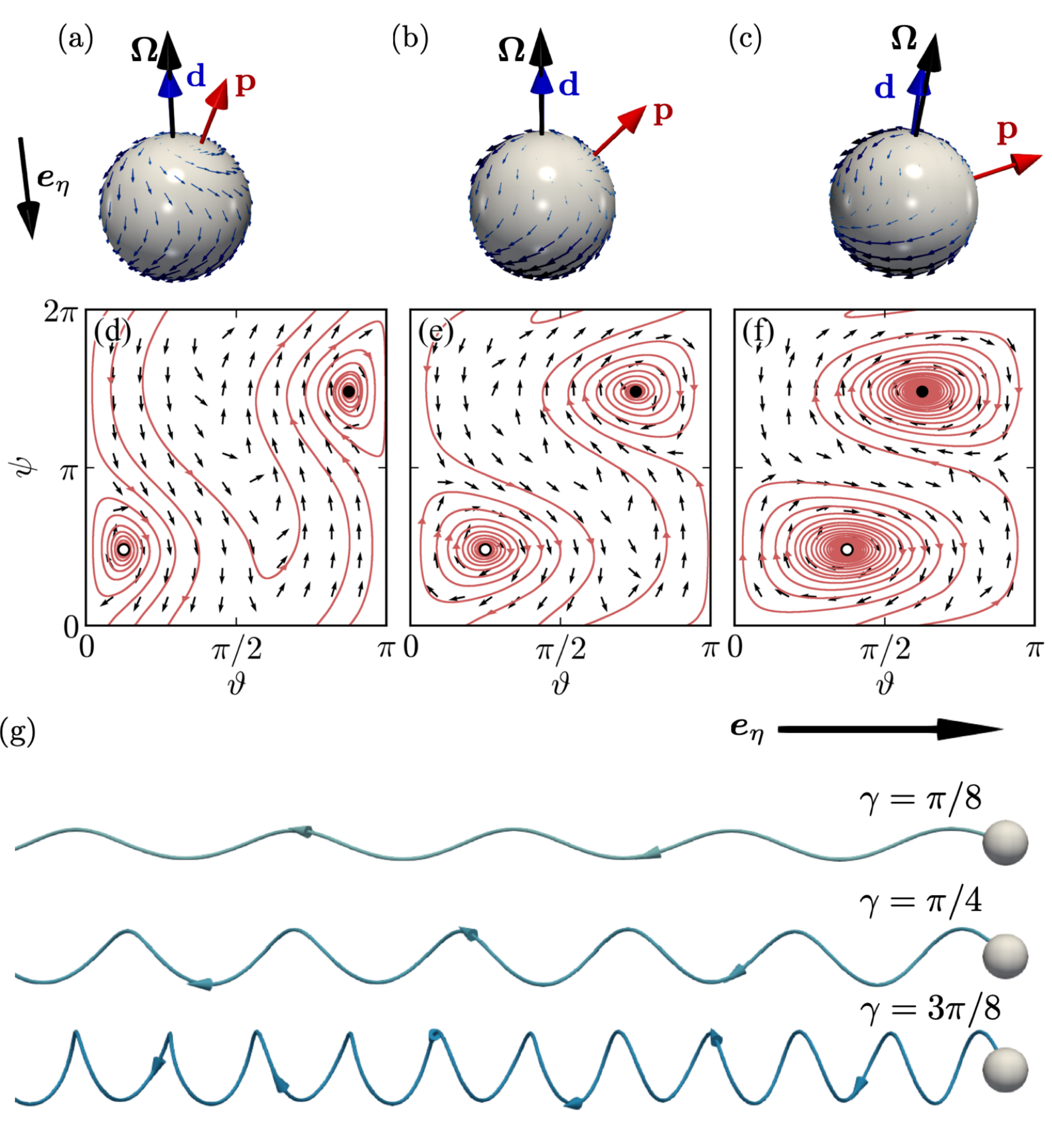}
    \caption{
    (a)-(c) Schematic representations of a chiral swimmer’s steady-state orientation for different angles: (a) $\gamma = \pi / 8$, (b) $\gamma = \pi / 4$, and (c) $\gamma = 3\pi / 8$, with fixed parameters $\varepsilon = 0.1$ and $\zeta = 5$.
    (d)-(f) Phase portraits of the orientational dynamics of a chiral squirmer for various angles: (d) $\gamma = \pi / 8$, (e) $\gamma = \pi / 4$, and (f) $\gamma = 3\pi / 8$, with $\varepsilon = 0.1$ and $\zeta = 5$. Each phase portrait shows one spiral source (white circle) and one spiral sink (black circle), with red lines representing example trajectories.
    (g) Steady-state trajectories of swimmers for different angles $\gamma$. All swimmers exhibit negative viscotaxis.
    }
    \label{fig4}
\end{figure}

In this Letter, we investigate the propulsion of a chiral microswimmer in a constant viscosity gradient using the canonical spherical squirmer model~\cite{Lighthill1952-ui, Blake1971-ws, Pak2014-uf} (\Figref{fig1}). We derive analytical expressions for the velocities of a single chiral microswimmer and demonstrate that the coupling between the swimmer’s chirality and viscosity gradients leads to intriguing orientation dynamics (\Figref{fig2}). These dynamics can be described by the Landau–Lifshitz–Gilbert (LLG) equations, commonly used for spin systems~\cite{Gilbert2004-gi, Lakshmanan2011-yl}.
Chiral microswimmers exhibit negative viscotaxis, moving toward regions of lower viscosity with spiral motions. Their trajectories transition from spiral (chiral) to straight (achiral) in viscosity gradients (\Figref{fig3}). Furthermore, we show that chiral microswimmers with varying angles between the source and rotlet dipoles always spiral toward lower viscosity regions, regardless of their initial orientation. This behavior contrasts with that of chiral swimmers with fixed dipole angles, whose trajectories depend on their initial conditions (\Figref{fig4}).
Our findings highlight the potential of leveraging distinct trajectories in viscosity gradients as a tool for cell sorting—a crucial method widely employed in biology and medical applications for isolating and characterizing target cells~\cite{Datt2019-jv, Tamura2014-qe}.

{\it Model}.---
We employ the squirmer model~\cite{Lighthill1952-ui, Blake1971-ws, Pak2014-uf} to represent microswimmers, where propulsion is generated by a slip velocity at the swimmer's surface.
We focus on the first polar mode $B_1$ (known as the source dipole) and the second azimuthal mode $C_2$ (known as the rotlet dipole), which captures the chirality. The resulting slip velocity at the surface is given by [see \Figref{fig1}(b, c)],
\begin{align}\label{surface_vel}
    \bm{u}^\mathcal{S}(\theta, \phi) = B_1\sin\theta\hat{\bm{\theta}} + \frac{3}{2}C_2\sin2\theta\hat{\bm{\phi}}.
\end{align}
where $\hat{\bm{\theta}}$ and $\hat{\bm{\phi}}$ are the unit polar and azimuthal tangent vectors, respectively.
The first polar mode $B_1$ determines the steady swimming speed in a uniform Stokes fluid, given by $U_{N} = (2/3)B_1$. The ratio $\zeta = C_2 / B_1$ indicates the strength of swirling (chiral) flow. The chiral parameter $\zeta$ describes the flow pattern of a microswimmer with rotating flagella and a counter-rotating body~\cite{Fadda2020-ve, Binagia2020-dn, Nganguia2020-ss, Housiadas2021-qe, Kobayashi2023-ad, Kobayashi2024-tn}.
The angular velocity in uniform Stokes fluids with constant viscosity is $\bm{\Omega} = \bm{0}$, meaning squirmers swim in straight lines.

We study a Newtonian fluid with viscosity variations influenced by factors such as temperature~\cite{Oppenheimer2016-bj} and nutrient concentration~\cite{Shoele2018-dj}. The fluid is incompressible and governed by the Stokes equations:
$\bm{\nabla}\cdot\bm{u} =0$ and $\bm{\nabla}\cdot\bm{\sigma} = \bm{0}$,
where $\bm{u}$ and $\bm{\sigma}$ represent the velocity and stress fields, respectively.
The stress field follows the Newtonian model but incorporates a spatially varying viscosity $\eta(\bm{r})$. Therefore, the governing equation of fluids is given by,
\begin{align}
    -\bm{\nabla}p + \eta(\bm{r})\bm{\nabla}^2\bm{u} + (\bm{\nabla}\eta)\cdot\left[\bm{\nabla}\bm{u} + (\bm{\nabla}\bm{u})^T\right] = \bm{0},
\end{align}
where $p$ is the pressure.
We consider a linear viscosity gradient, 
\begin{align}\label{viscosity}
    \bm{\nabla}\eta = \frac{\eta_0}{L} \bm{e}_\eta = \varepsilon \frac{\eta_0}{a}\bm{e}_\eta,\qquad \left(\varepsilon\equiv \frac{a}{L}\right),
\end{align} 
with the gradient directed along $\bm{e}_\eta$. 
While swimming motion can influence viscosity variations, we assume that fluid flow does not significantly affect the viscosity fields, meaning that the Péclet number, associated with the diffusion of the viscosity field, is small~\cite{Shoele2018-dj}. Therefore, the viscosity field remains as described by Eq.~\eqref{viscosity}, as assumed in previous studies~\cite{Liebchen2018-yb, Datt2019-jv, Shaik2021-ha, Gong2023-xt, Gong2024-no, Gong2024-yi}.
To facilitate an analytical treatment, we assume small viscosity variations~\cite{Oppenheimer2016-bj, Shoele2018-dj,Liebchen2018-yb, Datt2019-jv, Shaik2021-ha, Gong2023-xt, Gong2024-no, Gong2024-yi}, meaning the length scale $L$ significantly exceeds the swimmer's size $a$ ($\varepsilon \ll 1$).
We also ignore thermal fluctuations, focusing only on the deterministic motion of microswimmers~\cite{Liebchen2018-yb, Datt2019-jv, Shaik2021-ha, Gong2023-xt, Gong2024-no, Gong2024-yi}.

We solve the swimming problem using a regular perturbation expansion of $\varepsilon$, expressed as
\begin{align}
    \{\bm{u},\bm{\sigma}\} =\{\bm{u}_0,\bm{\sigma}_0\} + \varepsilon\{\bm{u}_1,\bm{\sigma}_1\}+\mathcal{O}(\varepsilon^2),
\end{align}
Applying this perturbation expansion of $\varepsilon$ to the integral theorem for Stokes fluids with viscosity gradients~\cite{Oppenheimer2016-bj, Shoele2018-dj}, we obtain [see Supplemental Material (SM) for detailed derivations],
\begin{align}\label{eq:theorem}
    -\int_\mathcal{S}\bm{n}\cdot\bm{\sigma}\cdot\hat{\bm{u}}\ dS + \int_\mathcal{S}\bm{n}\cdot\hat{\bm{\sigma}}\cdot\bm{u}\ dS = 2\varepsilon\int_\mathcal{V}\eta_1\bm{E}_0:\bm{\nabla}\hat{\bm{u}}\ dV,
\end{align}
where ${\bm{E}}_0$ is the strain-rate tensor corresponding to the zeroth-order (Stokes limit) velocity $\bm{u}_0$.

{\it Alignment of the Source and Rotlet dipoles}.---
To derive the swimming speeds of chiral squirmers, we use \Eqref{eq:theorem} and substitute the Stokes solutions for velocity and stress around a no-slip particle~\cite{Happel1983-ko} into the auxiliary problem, indicated with the hat~\cite{Stone1996-mx}. 
By applying the force-free and torque-free conditions for squirmers, we derive the swimming velocities [see Supplemental Material (SM)],
\begin{subequations}
    \begin{align}
        \bm{U} &=\bm{U}_0 = \frac{2}{3}B_1\mathbf{p},\\
        \bm{\Omega} &=-\frac{B_1}{3}\mathbf{p}\times \bm{\nabla}\left(\frac{\eta}{\eta_0}\right) + \frac{2}{5}C_2(\bm{1}-3\mathbf{pp})\cdot\bm{\nabla}\left(\frac{\eta}{\eta_0}\right).\label{eq:omega2}
    \end{align}
\end{subequations}
The viscosity gradient only affects the angular velocity of the chiral squirmer with $B_1$ and $C_2$ modes. The first term of Eq.~\eqref{eq:omega2} indicates that $B_1$ mode causes the squirmer to rotate toward regions of lower viscosity (negative viscotaxis) when not perfectly aligned with the viscosity gradient~\cite{Datt2019-jv}. 
The angular velocity due to the chirality in the linear viscosity gradient can be decomposed by
\begin{gather}\label{eq:omega}
    \begin{split}
        \bm{\Omega} = \Omega_{\rm Sp} (\bm{e}_\eta\cdot\mathbf{p})\mathbf{p} + \Omega_{\rm Pr}\bm{e}_\eta + \Omega_{\rm Da}\bm{e}_\eta \times \mathbf{p},\\
        \Omega_{\rm Sp} = -\frac{6C_2}{5L},\quad
        \Omega_{\rm Pr} = \frac{2C_2}{5L},\quad
        \Omega_{\rm Da} = \frac{B_1}{3L}.
    \end{split}
\end{gather}
Using Eq.~\eqref{eq:omega} and $\dot{\mathbf{p}} = \bm{\Omega}\times\mathbf{p}$, we derive the orientational dynamics of the chiral squirmer [see \Figref{fig2}(a)], governed by the Landau--Lifshitz--Gilbert (LLG) equation, capturing the essential physics of the alignment behavior~\cite{Gilbert2004-gi, Lakshmanan2011-yl}
\begin{align}\label{LLG}
    \dot{\mathbf{p}} = \Omega_{\rm Pr}(\bm{e}_\eta\times\mathbf{p}) + \Omega_{\rm Da}(\bm{e}_\eta\times\mathbf{p})\times\mathbf{p}.
\end{align}
The swimmer exhibits damping ($\Omega_{\rm Da}$) leading to negative viscotaxis, spinning ($\Omega_{\rm Sp}$) around the propulsion axis $\mathbf{p}$ and precession ($\Omega_{\rm Pr}$) around the axis of viscosity gradients $\bm{e}_\eta$.
Interestingly, the orientation dynamics of achiral squirmers in chiral fluids with odd viscosity are governed by the fixed LLG eq., exhibiting parallel or antiparallel alignments with the axis of odd viscosity~\cite{Hosaka2024-zy}. 
The interplay between a polar vector (orientation) and a pseudovector (axis of odd viscosity) makes parallel and antiparallel alignments equivalent.
In spin systems, however, parallel and antiparallel alignments are not equivalent, as both magnetization (or spin) and the magnetic field are pseudovectors. 
In viscosity gradients, both the orientation $\mathbf{p}$ and $\bm{e}_\eta$ are polar vectors, rendering parallel and antiparallel alignments nonequivalent. Consequently, the orientation dynamics can be elegantly described by LLG eq.
Moreover, while experiments on microswimmers in fluids with odd viscosity remain elusive, numerous studies on viscosity gradients~\cite{Kaiser1975-nl, Petrino1978-wh, Takabe2017-tp, Daniels1980-ue, Sherman1982-ur, Stehnach2021-nv, Coppola2021-iu} suggest our findings are experimentally reproducible.

The intriguing orientation dynamics can be understood by examining the effect of the viscosity gradient on the thrust torque. The thrust torque, generated by the slip velocity at the squirmer's surface, dominates in regions of higher viscosity. As shown in \Figref{fig2}(b) and (c), when the swimming axis is perpendicular (antiparallel) to the viscosity gradient, the torque is predominantly governed by the traction generated by the right (lower) hemisphere, leading to rotating around $\bm{e}_\eta$ ($\mathbf{p}$). When not perfectly aligned with the viscosity gradient, chiral squirmers move toward regions of lower viscosity, following spiral paths. Once aligned with the viscosity gradient, the squirmer swims in a straight line with spinning. 
Chiral squirmers typically transition from chiral (spiral) to achiral trajectories (linear trajectories with spinning) unless perfectly aligned with viscosity gradients. 

Now, we describe the swimmer's rotation using the polar angle $\vartheta$, measured from the $\bm{e}_\eta$-axis, meaning $\vartheta = \cos^{-1}(\mathbf{p}\cdot\bm{e}_\eta)$, and the azimuthal angle $\varphi$. The swimming axis is $\mathbf{p} = (\sin\vartheta\cos\varphi, \sin\vartheta\sin\varphi, \cos\vartheta)$, and \Eqref{LLG} becomes
\begin{align}\label{eq:thetaphi}
    \dot{\vartheta} = \Omega_{\rm Da}\sin\vartheta,\qquad
    \dot{\varphi} = \Omega_{\rm Pr}.
\end{align}
The dynamics of $\vartheta$ are characterized by one unstable fixed point at $\vartheta = 0$ (corresponding to positive viscotaxis) and one stable fixed point at $\vartheta = \pi$ (corresponding to negative viscotaxis).
Equations.~\eqref{eq:thetaphi} solve to
\begin{align}
    \tan\frac{\vartheta(t)}{2} = \tan\frac{\vartheta(0)}{2}e^{\Omega_{\rm Da}t},\quad
    \varphi(t) = \Omega_{\rm Pr}t + \varphi(0).
\end{align}
We demonstrate negative viscotaxis, accompanied by spiral trajectories, for various chiral parameters $\zeta = C_2 / B_1$ and initial polar angles $\vartheta(0)$ in~\Figref{fig3}. Chirality allows the swimmer to follow a three-dimensional trajectory, in contrast to an achiral swimmer, which moves within a two-dimensional plane~\cite{Datt2019-jv}.
As shown in~\Figref{fig3}, the specific trajectories vary depending on the swimmer's chirality, orientation, and viscosity gradient strengths. When the swimmer's orientation is parallel or antiparallel to the axis of viscosity gradients $\bm{e}_\eta$, it swims in a straight line. In contrast, all chiral squirmers follow spiral trajectories and exhibit negative viscotaxis unless perfectly aligned with the axis of viscosity gradients. Eventually, the swimmer will align perfectly with $-\bm{e}_\eta$ and exhibit straight-line motion. Particularly, the angular velocity associated with precession $\Omega_{\rm Pr}$ increases with the chiral parameter $\zeta$, resulting in smaller radii for the spiral trajectories.
Additionally, stronger viscosity gradients lead to larger angular velocities associated with both precession and damping, resulting in tighter spiral trajectories. 
These distinctions in trajectories could potentially be exploited for sorting swimmers~\cite{Tamura2014-qe}. Furthermore, the swimmer's paths could be used to measure the fluid's viscosity gradient strengths.

{\it Misalignment of the Source and Rotlet Dipoles}.---
We next investigate a chiral swimmer with a rotlet dipole oriented in the $\mathbf{d}$ direction, fixed within the body frame and forming an angle $\gamma$ with the swimming axis $\mathbf{p}$. This configuration is reminiscent of the tumbling behavior observed in {\it E.coli} cells. When all flagella of {\it E.coli} rotate in the same direction, the cells swim in straight lines. However, when at least one flagellum rotates in the opposite direction, the cells exhibit a tumbling motion~\cite{Macnab1977-ri}.
We solve for the velocities in linear viscosity fields up to $\mathcal{O}(\varepsilon)$,
\begin{align}\label{eq:omega3}
    \bm{\Omega} = \Omega_{\rm Sp}(\bm{e}_\eta\cdot\mathbf{d})\mathbf{d} + \Omega_{\rm Pr}\bm{e}_\eta + \Omega_{\rm Da}\bm{e}_\eta\times\mathbf{p}.
\end{align}

We represent the swimmer's rotation using the Euler angles $(\varphi, \vartheta, \psi)$, where $\varphi$ and $\vartheta$ are the azimuthal and polar angles measured from the $\bm{e}_z = \bm{e}_\eta$ axis, and $\psi$ is the roll angle about the direction $\mathbf{p}$.
By representing $\mathbf{d} = \sin\gamma\bm{e}_1 + \cos\gamma\bm{e}_3$ without loss of generality, and substituting into~\Eqref{eq:omega3}, we obtain [see Supplemental Material (SM)]
\begin{subequations}
    \begin{align}
        \dot{\varphi} &= \Omega_{\rm Sp}\sin\gamma\sin\psi\left(\sin\gamma\sin\psi + \frac{\cos\gamma}{\tan\vartheta}\right) +  \Omega_{\rm Pr},\\
        \begin{split}
            \dot{\vartheta} &= \Omega_{\rm Sp}\sin\gamma\cos\psi(\sin\gamma\sin\vartheta\sin\psi + \cos\gamma\cos\vartheta)\\
            &\quad+ \Omega_{\rm Da}\sin\vartheta,
        \end{split}\\
        \begin{split}
            \dot{\psi} &= \Omega_{\rm Sp}\Bigg(\cos^2\gamma\cos\vartheta - \sin^2\gamma\cos\vartheta\sin^2\psi \\
            &\qquad+ \sin\gamma\cos\gamma\frac{\sin^2\vartheta- \cos^2\vartheta}{\sin\vartheta}\sin\psi\Bigg).
        \end{split}
    \end{align}
\end{subequations}
Since the equations are independent of $\varphi$, it suffices to examine the phase portraits in the $(\vartheta,\psi)$ plane [see~\Figref{fig4}(d)-(f)]. The system displays a spiral source and sink, with the sink indicating negative viscotaxis. At the steady state, the axis of angular velocity $\bm{\Omega}$ is nearly antiparallel to the axis of viscosity gradients [see~\Figref{fig4}(a)-(c)]. Consequently, the steady-state trajectories of chiral squirmers are spiral, with the radius of these spiral trajectories decreasing with the angle $\gamma$ [see~\Figref{fig4}(g)]. 
Swimmers with $\mathbf{d} \ne \mathbf{p}$ consistently exhibit spiral trajectories, unlike swimmers with $\mathbf{d} = \mathbf{p}$. Furthermore, the unstable point, which corresponds to the positive viscotaxis for swimmers with $\mathbf{d}=\mathbf{p}$, disappears for swimmers with $\mathbf{d}\ne\mathbf{p}$. 
Additionally, increasing the chiral parameters $\zeta$ results in smaller radius, while increasing the viscosity gradients $\varepsilon$ leads to tighter spiral trajectories.

{\it Conclusion.}--- 
We present analytical expressions for the velocities of chiral microswimmers in viscosity gradients. The interaction between the swimmer’s chirality and the fluid’s viscosity gradient induces an additional rotation described by the Landau–Lifshitz–Gilbert (LLG) equations. Squirmers exhibit three key dynamics: damping ($\Omega_{\rm Da}$), associated with negative viscotaxis; spinning ($\Omega_{\rm Sp}$) around the propulsion axis $\mathbf{p}$; and precession ($\Omega_{\rm Pr}$) around the viscosity gradient axis $\bm{e}_\eta$. Similar orientation behaviors are observed in achiral swimmers within chiral fluids, such as those with odd viscosity~\cite{Hosaka2024-zy}, indicating that both swimmer and fluid chirality can generate chiral motion. Notably, chiral squirmers with misaligned dipoles ($\mathbf{d}\ne\mathbf{p}$) follow steady-state spiral trajectories aligned with negative viscotaxis.

Future studies will explore not only single-swimmer phenomena, such as traction-driven microswimmers~\cite{Ishikawa2024-vp, Hosaka2023-tt, Kobayashi2024-ke} and viscoelastic effects (durotaxis)~\cite{Shaik2024-eq}, but also collective behaviors in viscosity gradients~\cite{Samatas2023-dg}. The observed spiral motion of charged particles in constant magnetic fields~\cite{Lluis_Gonzalez2022-lx} may provide further analogies to our findings. Similarity to the Vicsek model~\cite{Vicsek1995-zy}, where the orientations are governed by the XY model for spin systems, our findings on orientation dynamics (LLG eq.) offer valuable insights into the connections between active matter and spin systems.
The transition mechanisms we identify offer insights into the diverse swimming behaviors influenced by fluid rheology, such as E. coli’s run-and-tumble motion~\cite{Kurzthaler2024-ip} or mammalian sperm chemokinetics~\cite{Zaferani2023-yy}. Finally, our results could inform the development of tools~\cite{Tamura2014-qe} for cell sorting and quantifying viscosity gradients in complex fluids.

\begin{acknowledgments}
We thank Norihiro Oyama for his careful reading of the manuscript and constructive comments. We also thank Raymond E. Goldstein, Ronojoy Adhikari, Mitsusuke Tarama, and Yuto Hosaka for fruitful discussions.
This work was supported by the Grants-in-Aid for Scientific Research (JSPS KAKENHI) under Grant Number 20H05619, by the JSPS Core-to-Core Program ``Advanced core-to-core network for the physics of self-organizing active matter (JPJSCCA20230002)'' and by JST SPRING, Grant Number JPMJSP2110.
\end{acknowledgments}

%

\end{document}